\documentclass[prd,preprintnumbers,superscriptaddress,floatfix,eqsecnum,nofootinbib,12pt]{revtex4}

\usepackage[T1]{fontenc}
\usepackage[latin1]{inputenc}
\usepackage[english]{babel}
\usepackage{amssymb,amsmath,amsthm,amsfonts}
\usepackage{hyperref}
\usepackage{cleveref}
\pdfoutput=1
\usepackage[]{graphicx}
\usepackage[]{subfigure}
\usepackage[usenames,svgnames]{xcolor}
\usepackage[letterpaper,left=2.cm,right=2.cm,top=2.5cm,bottom=2.5cm]{geometry}
\usepackage{tensor}
\usepackage{lmodern}
\usepackage[free-standing-units, exponent-product = \cdot, list-units = single, range-units = single]{siunitx}
\usepackage{csquotes}

\usepackage{lipsum}
\usepackage{bibentry} 

\usepackage{tikz}
\usepackage{pgfplots}

\usetikzlibrary{external}
\usetikzlibrary{calc}
\usepgfplotslibrary{groupplots}

\pgfplotsset{
    grid = both,
    grid style = {opacity = 0.5},
    compat = newest,
    every axis plot/.append style={
        line join=round,
        clip=false
    },
}

\definecolor{colorzero}{HTML}{000000}
\definecolor{colorone}{HTML}{E41A1C}
\definecolor{colortwo}{HTML}{FF7F00}
\definecolor{colorthree}{HTML}{4DAF4A}
\definecolor{colorfour}{HTML}{377EB8}
\definecolor{colorfive}{HTML}{984EA3}

\newcommand{\ie}{\textit{i.e}}
\newcommand{\eg}{\textit{e.g}}
\newcommand{\dd}{\mathrm{d}}

\usepackage{acro}
\DeclareAcronym{gr}{
    short=GR ,
    long=general relativity
}
\DeclareAcronym{uv}{
    short=UV ,
    long=ultraviolet
}
\DeclareAcronym{ir}{
    short=IR ,
    long=infrared
}
\DeclareAcronym{wdw}{
    short=WDW ,
    long=Wheeler-DeWitt
}
\DeclareAcronym{hl}{
    short=HL ,
    long=Ho\v{r}ava-Lifshitz
}
\DeclareAcronym{rg}{
    short=RG ,
    long=renormalization group
}
\DeclareAcronym{adm}{
    short=ADM ,
    long={Arnowitt, Deser and Misner}
}

\begin{document}

\preprint{YITP-22-52, IPMU22-0028}

\title{DeWitt wave function in Ho\v{r}ava-Lifshitz cosmology\\
 with tensor perturbation}

\author{Paul Martens}
\email[]{paul.martens@yukawa.kyoto-u.ac.jp}
\affiliation{Center for Gravitational Physics and Quantum Information, Yukawa Institute for Theoretical Physics, Kyoto University, 606-8502, Kyoto, Japan}
\author{Hiroki Matsui}
\email[]{hiroki.matsui@yukawa.kyoto-u.ac.jp}
\affiliation{Center for Gravitational Physics and Quantum Information, Yukawa Institute for Theoretical Physics, Kyoto University, 606-8502, Kyoto, Japan}
\author{Shinji Mukohyama}
\email[]{shinji.mukohyama@yukawa.kyoto-u.ac.jp}
\affiliation{Center for Gravitational Physics and Quantum Information, Yukawa Institute for Theoretical Physics, Kyoto University, 606-8502, Kyoto, Japan}
\affiliation{Kavli Institute for the Physics and Mathematics of the Universe (WPI), The University of Tokyo Institutes for Advanced Study, The University of Tokyo, Kashiwa, Chiba 277-8583, Japan}

\date{\today}

\begin{abstract}
We present a well-tempered DeWitt wave function, which vanishes at the classical big-bang singularity, in Ho\v{r}ava-Lifshitz (HL) cosmology with tensor perturbation, both analytically and numerically. In general relativity, the DeWitt wave function is ill-behaved once the tensor perturbation is taken into account. This is essentially because the amplitude of the perturbation diverges at the singularity and the perturbative expansion completely breaks down. On the other hand, in HL gravity it is known that the higher dimensional operators required by the perturbative renormalizability render the tensor perturbation scale-invariant and regular all the way up to the singularity. In this paper we analytically show that in $d+1$ dimensional HL gravity the DeWitt wave function for tensor perturbation is indeed well-defined around the classical big-bang singularity. Also, we numerically demonstrate the well-behaved DeWitt wave function for tensor perturbation from the big-bang to a finite size of the Universe. 
\end{abstract}

\maketitle
\newpage
\tableofcontents

\section{Introduction}
\label{sec:intro} 

Attempts to build a quantum field theory of gravity have encountered many serious issues. As is well known, \ac{gr} is not renormalizable, essentially because Newton's constant is dimensionfull (with the mass dimension $\left[G_{N}\right]=-2$). The non-renormalizability leads to uncontrollable \ac{uv} divergences. Adding higher curvature terms to the Einstein-Hilbert action can make the theory renormalizable~\cite{Stelle:1976gc}, or even super-renormalizable~\cite{Asorey:1996hz}, but leads to massive ghosts and makes the corresponding quantum theory non-unitary in the UV. 

That is where \ac{hl} gravity~\cite{Horava:2009uw} enters. This quantum gravity theory is power-counting renormalizable, thanks to the presence of higher-order spatial curvature terms. The action and equations of motion in \ac{hl} gravity contain only terms up to second order in time derivatives and thus the theory is free from Ostrogradsky ghosts associated with higher time derivatives. The \ac{hl} theory was recently proved to be perturbatively renormalizable~\cite{Barvinsky:2015kil,Barvinsky:2017zlx}, and, therefore, has come to be regarded as a valid \ac{uv} completion path of quantum gravity.

One of the fundamental principles of the \ac{hl} gravity is the so-called \textit{anisotropic scaling}, or \textit{Lifshitz scaling}, \ie.
\begin{equation}
 t \to b^z t\,, \quad \vec{x} \to b \vec{x}\,, 
\end{equation}
where $t$ is the time coordinate, $\vec{x}$ represents the spatial coordinates vector and $z$ is a number called \textit{dynamical critical exponent}. In $3+1$ dimensions, an anisotropic scaling of $z=3$ in the \ac{uv} regime breaks Lorentz symmetry, but ensures renormalizability, while the usual $z=1$ scaling is recovered in the \ac{ir} regime. An anisotropic scaling of $z=3$ also handily provides a mechanism for generating scale-invariant cosmological perturbations, simultaneously solving the horizon problem~\cite{Mukohyama:2009gg} without inflation and providing the so-called \textit{anisotropic instanton}, which could be an answer to the flatness problem~\cite{Bramberger:2017tid}. 

Although there have been numerous works on the early cosmology based on the HL gravity, the early Universe cannot be treated classically when quantum gravity is dominant, and we have to adopt a \textit{quantum cosmology} approach. A common approach, in quantum cosmology, is to separate space and time using the \ac{adm} formalism and work with the Hamiltonian formulation. In this framework, the Hamiltonian constraint is interpreted as an operator equation, the \ac{wdw} equation~\cite{DeWitt:1967yk}, $\hat{H}[g]\Psi[g]=0$ where $\hat{H}[g]$ is the Hamiltonian operator and $\Psi[g]$ is called the \textit{wave function of the Universe} (see \cite{Halliwell:1989myn,Wiltshire:1995vk} for a review).

Whether this wave function of the Universe gives finite correlation functions of physical perturbations has a long history of debate among quantum cosmologists. In particular, in recent years, in the Lorentzian path integral formulation of the no-boundary~\cite{Hartle:1983ai} and tunneling proposals~\cite{Vilenkin:1984wp}\,---well-known boundary conditions of quantum cosmology---, it has been argued that the wave function of small perturbations around the background may take an inverse Gaussian form and become uncontrollable~\cite{Feldbrugge:2017fcc,Feldbrugge:2017mbc}. In this work, we adopt instead the so-called the DeWitt boundary condition~\cite{DeWitt:1967yk}, which states that the wave function of the Universe should vanish at the classical big-bang singularity. In a homogeneous and isotropic Universe, the DeWitt boundary condition can be concisely expressed as $\Psi(a=0)=0$ and is known to successfully regularize the behavior of the wave function near the classical singularity. In a mini-superspace, where the dynamics of the Universe is parameterized only by the scale factor $a(t)$, it is easy to find an analytic expression for the \textit{DeWitt wave function}, \ie. a solution to the \ac{wdw} equation with the DeWitt boundary condition. However, how to generalize this result beyond this mini-superspace is not obvious. 

In a previous work~\cite{Matsui:2021yte}, we have analytically shown that, in GR and many other theories of gravity, the DeWitt wave function for a homogeneous and isotropic background with tensor perturbations is not well-behaved. The supposedly small perturbations are not suppressed near the classical big-bang singularity. This is a serious problem if the DeWitt wave function is adopted as a description of the very early Universe in theories of gravity, including \ac{gr}. Therefore, the DeWitt boundary condition, proposed to overcome the classical singularity in quantum cosmology, requires a description of gravity that goes beyond \ac{gr}. Fortunately, in the \ac{hl} gravity, the wave function of the tensor perturbation is found to behave well, thanks to the presence of higher dimensional operators that are required for perturbative renormalizability. 

We here extend this result from $3+1$ dimensions to $d+1$ dimensions, and show that the DeWitt wave function for tensor perturbation exhibits scale invariance near the classical big-bang singularity and thus is regular. We also demonstrate the well-behaved DeWitt wave function for tensor perturbation numerically. This numerical demonstration for the first time illustrates the regular behavior of the DeWitt wave function for tensor perturbation in \ac{hl} gravity all the way from the classical big-bang singularity to finite values of the scale factor. 

The rest of the present paper is organized as follows: in \cref{sec:formulation} we give a brief review of the quantum cosmology framework in the $d + 1$ dimensional HL gravity. In \cref{sec:with-tensor}, we derive the DeWitt wave function for a $d+1$ dimensional Universe described by a homogeneous and isotropic background and tensor perturbation around it, and show that the wave function for the tensor perturbation is normalizable and scale-invariant on constant-$a$ hypersurfaces near $a=0$, where $a$ is the scale factor. This result is due to the anisotropic scaling with $z=d$. Numerical results are demonstrated for $d=3$ for concreteness. Finally, we conclude our work in \cref{sec:conclusion}.

\section{Framework}
\label{sec:formulation} 

\subsection{Ho\v{r}ava-Lifshitz gravity}

In this section we will briefly review the construction of \ac{hl} gravity in $d + 1$ dimensions and introduce the \ac{wdw} equation. Similar to the \ac{adm} decomposition of the metric in \ac{gr}~\cite{Arnowitt:1962hi}, we can write the $d + 1$ dimensional metric as
\begin{align}
\dd s^2 = - N^2 \dd t^2 + g_{i j} (\dd x^i + N^i \dd t) (\dd x^j + N^j \dd t) \,,
\end{align}
where $i=1,\dots, d$, $N$ is the lapse function, $N ^i$ the shift vector, and $g_{ij}$ is the spatial metric with positive definite signature $(+,+,\cdots)$. The $d + 1$ dimensional action $S$ describing \ac{hl} gravity is the sum of a kinetic part $\mathcal{L}_K$ and a potential part $\mathcal{L}_V$~\cite{Mukohyama:2010xz}; that is
\begin{equation}
\label{action}
    S= \frac{{\cal M}_{\rm HL}^{d-1}}{2}\int \dd t \dd^d x N \sqrt{g} \left( {\cal L}_{K}+{\cal L}_{V}\right)\,, 
\end{equation}
where ${\cal M}_{\rm HL}$ sets the overall mass scale. The kinetic part $\mathcal{L}_K$ can be expressed in terms of the extrinsic curvature $K_{ij}$ as
\begin{equation}
    \label{curvature1}
    {\cal L}_{K}= (K^{ij}K_{ij}-\lambda K^2)\,, 
    \quad\text{where}\quad
    K_{ij}=\frac{1}{2N} (\partial_t g_{ij}- g_{j k}\nabla_iN^k-g_{i k}\nabla_jN^k)\,,
\end{equation}
with $\lambda$ being a coupling constant, and $\nabla_i$ the spatial covariant derivative compatible with $g_{ij}$. The indices $i,j$ run from $1$ to $d$ and the inverse of $g_{ij}$ is written as $g^{ij}$. Accordingly, we wrote $K^{ij}=g^{ik}g^{jl}K_{kl}$ and $K=g^{ij}K_{ij}$. For simplicity, we will consider the projectable \ac{hl} gravity where the lapse function depends only on time, \ie. $N=N(t)$. The results of the present paper can be directly applied to the non-projectable theory as well, by simply setting $C_{\alpha}=0$ (and thus $\mathcal{C}=0$).

As for the potential part, it shall be built out of any possible renormalizable operators without parity violating terms, and must be expressed in terms of invariants assembled from the products of the Riemann tensor (schematically denoted as Rm below) and its derivatives~\cite{Sotiriou:2009gy,Sotiriou:2009bx}. For the highest-order potential part, denoted as $\mathcal{L}_z$, the list of possible terms is 
\begin{equation}
\{ (\textrm{Rm})^d,\, (\nabla\textrm{Rm})^2(\textrm{Rm})^{d-3},\,
 (\nabla\textrm{Rm})^4(\textrm{Rm})^{d-6},\, (\nabla\textrm{Rm})^6(\textrm{Rm})^{d-9},
 \cdots\} \, .
\end{equation}

Now, including lower-dimensions terms as well, we come to write
\begin{align}
    \begin{split}
        {\cal L}_{V} & = {\cal L}_{0}+{\cal L}_{1}+{\cal L}_{2}+{\cal L}_{3}+\cdots +{\cal L}_{z-1}+{\cal L}_{z} \, ,
    \end{split}
\end{align}
where ${\cal L}_i$ denotes a collection of invariant terms with $2i$ spatial derivatives acted on the spatial metric. Here, the first two terms ${\cal L}_{0}+{\cal L}_{1}$ are necessary to recover the $d + 1$-dimensional \ac{gr} in the \ac{ir} limit, the third term ${\cal L}_{2}$ includes all possible quadratic and spatial curvature terms, and so on. Explicitly, the four five terms $\mathcal{L}_i$ read 
\begin{subequations}
    \begin{align}
        {\cal L}_{0} & = -2\Lambda \,, \\
        {\cal L}_{1} & = c^2_g R \,, \\
        {\cal L}_{2} & = c_{2,1}R^2+c_{2,2}R_i^jR_j^i+c_{2,3}R^{ijkl}R_{ijkl} \, , \\
        {\cal L}_{3} & = c_{3,1}R^3+c_{3,2}RR_i^jR_j^i+c_{3,3}R_i^jR_j^kR_k^i+c_{3,4}\nabla_iR\nabla^iR +c_{3,5}\nabla_iR_{jk}\nabla^iR^{jk} + \cdots \, , \\
        \begin{split}
            {\cal L}_{4} & = c_{4,1}R^4 + c_{4,2}R^2R_i^jR_j^i + c_{4,3}RR_i^jR_j^kR_k^i+c_{4,4}R_i^jR_j^kR_k^lR_l^i \\
            & \phantom{= c_{4,1}R^4 + c_{4,2}R^2R_i^jR_j^i} +c_{4,5}R\,\nabla_iR\nabla^iR +c_{4,6}R\,\nabla_iR_{jk}\nabla^iR^{jk}+\cdots \, ,
        \end{split}
    \end{align}
\end{subequations}
where $R$, $R_{ij}$ and $R_{ijkl}$ respectively denote the Ricci scalar, the Ricci tensor and the Riemann tensor, and dots represent terms depending on the Riemann tensor and its derivatives. The constants $\Lambda$ and $c_g$ are the cosmological constant and the propagation speed of tensor gravitational waves, and $c_{m,n}$ are constants of appropriate dimensions. All those constants are subject to running under the \ac{rg} flow.

In the \ac{uv} regime, terms with two time derivatives and those with $2z$ spatial derivatives are dominant. If $z=d$, the theory is renormalizable and if $z>d$, the theory is super-renormalizable.
In the \ac{ir} regime, on the other hand, higher derivative terms are subdominant and the theory naturally converges to $z=1$. Moreover, if $\lambda$ (of \cref{curvature1}) goes to $1$ in the \ac{ir} limit, and if it does sufficiently quickly, then \ac{gr} is recovered thanks to an analogue of the Vainshtein mechanism~\cite{Mukohyama:2010xz,Izumi:2011eh,Gumrukcuoglu:2011ef}. For simplicity, in the rest of the present paper, we restrict our considerations to the renormalizable theory, \ie. $z=d$.

\subsection{Cosmological setup}

We shall assume the $d$-dimensional space of the model to be the union of connected pieces $\Sigma_{\alpha}$ ($\alpha=1,\dots$), each of which we shall call a \textit{local $d$-dimensional Universe}. The union of all $\Sigma_{\alpha}$ would thus represent the Universe in its entirety. In such a configuration, while the lapse function $N=N(t)$ is common for all pieces, we have a set of a shift vector $N^i = N_{\alpha}^i(t,x)$ and a spatial metric $g_{ij} = g^{\alpha}_{ij}(t,x)$, which is different for different local Universe $\Sigma_{\alpha}$. 
In the ADM formalism, the lapse and the shift vectors basically act as Lagrange multipliers. The variation of the projectable \ac{hl} action~(\cref{action}) with respect to $N$ yields the following \textit{Hamiltonian constraint} 
\begin{equation}
    \label{eq:Hamiltonian-constraint}
    \sum_{\alpha} \int_{\Sigma_{\alpha}}\dd^dx\, \mathcal{H}_{g\perp}=0 \, ,
    \quad \text{where} \quad 
    \mathcal{H}_{g\perp} = \frac{{\cal M}_{\rm HL}^{d-1}}{2}\sqrt{g}({\cal L}_{K}-{\cal L}_{V})\,.
\end{equation}
While this constrain sets the sum of contributions from all $\Sigma_{\alpha}$ to vanish, each contribution does not have to vanish, \ie.
\begin{equation}
    \int_{\{\Sigma_{\alpha}\}}d^dx\, \mathcal{H}_{g\perp} \ne 0 \, .
\end{equation}
Therefore, if we are interested in one particular local Universe, \ie. an element of $\{\Sigma_{\alpha}\}$, the Hamiltonian constraint does not need to be enforced.

Hereafter, we further assume that each connected space $\Sigma_{\alpha}$ is a closed universe, and that each closed universe is described by a closed Friedmann-Lema\^{i}tre-Robertson-Walker (FLRW) metric and perturbations around it. 
\begin{equation}
    N_{\alpha}^i = 0\,, \quad g^{\alpha}_{ij} = a^2_{\alpha}(t) \left[\Omega_{ij} ({\bf x}) +h_{i j}^{\alpha} (t \,, {\bf x}) \right]\,,
\end{equation}
where $\Omega_{ij}$ is the metric of the unit $d$-sphere with the curvature constant set to $1$, \ie. the Riemann curvature of $\Omega_ij$ is simply $\delta^i_k\delta^j_l-\delta^i_l\delta^j_k$. Given this definition, the spatial indices $i, j, \dots$ are thus raised and lowered by $\Omega^{ij}$ and $\Omega_{ij}$ respectively. The tensor perturbation $h^{\alpha}_{ij}$ must satisfy the transverse and traceless condition, \ie. $\Omega ^{i j} h^{\alpha}_{ij} = \Omega^{ki}D_k h^{\alpha}_{ij} = 0$, where $D_i$ is the spatial covariant derivative compatible with $\Omega_{ij}$. Furthermore, the perturbation $h_{ij}^{\alpha}$ can be expanded in terms of the hyper-spherical harmonics~\cite{Gerlach:1978gy} as
\begin{equation}\label{eq:perturbation}
    h_{ij}^{\alpha}(t,x^i)= \sum_{snlm} h^{\alpha s}_{nlm}(t)Q^{snlm}_{ij}\,,
\end{equation}
where $s=\pm$ is the polarization label and the triplet $(n, l, m)$ is comprised within the ranges $n \geq 3$, $l \in [0,n-1]$ and $m \in [-l,l]$. The tensor eigenfunctions $Q^{snlm}_{ij}$ of the Laplacian operator $D^2 [\Omega]$ on the unit $d$-sphere~\cite{Higuchi:1986wu} satisfy
\begin{equation}
    \label{eq:harmonics}
    D^2 [\Omega]\, Q^{snlm}_{ij}= -\left[n^2+(d-3)n-d\right]Q^{snlm}_{ij} \, ,
\end{equation}
and are normalized via the following relation
\begin{equation}
    \int d^d\sqrt{\Omega}\Omega^{ik}\Omega^{jl}Q^{snlm}_{ij}Q^{s'n'l'm'}_{kl} 
    = V_d\,\delta^{ss'}\delta^{nn'}\delta^{ll'}\delta^{mm'}\,. 
\end{equation}

With these assumptions and the harmonic expansion~\eqref{eq:perturbation}, the $d + 1$-dimensional \ac{hl} action is expanded up to the second order in perturbations as $S = S^{(0)}+S^{(2)}+\mathcal{O}(h^3)$, where
\begin{subequations}
    \label{eq:HL3}
    \begin{align}
        S^{(0)} &=  {\cal V} \sum_{\alpha}\int \dd t\, 
        \left({Na_{\alpha}^d}\right) \left[
              \frac{d(1-d\lambda)}{2}\left(\frac{\dot{a}_{\alpha}}{Na_{\alpha}}\right)^{2}
              + \frac{\alpha_d}{a^{2d}_{\alpha}}\cdots
              + \frac{\alpha_3}{a^6_{\alpha}}+\frac{\alpha_2}{a^4_{\alpha}}+c_{\rm g}^2
        \frac{d(d-1)}{2a^2_{\alpha}}-\Lambda \right]\,, \\
        S^{(2)} &=  {\cal V} \sum_{\alpha} \int \dd t \left({Na_{\alpha}^{d-2}}\right) 
        \sum_{snlm}
        \left[ \frac{1}{8}\left(\frac{a_{\alpha}}{N}\dot{h}^{\alpha s}_{nlm}\right)^2+ \left(\beta_1
        +\frac{\beta_2}{a^2_{\alpha}}+\frac{\beta_3}{a^4_{\alpha}}\cdots
        +\frac{\beta_d}{a^{2d-2}_{\alpha}}
        \right)\left(h^{\alpha s}_{nlm}\right)^2\right]\,. 
    \end{align}
\end{subequations}
Here, $\alpha_i$ and $\beta_i$ are constants that are linear combinations of coupling constants in the action, ${\cal V}={\cal M}_{\rm HL}^{d-1}V_d$ and $V_d=\int \dd^d x\sqrt{\Omega}$ is the volume of the unit $d$-sphere. For the tensor perturbation of each connected space $\Sigma_{\alpha}$, in the following for simplicity we restrict our attention to the dynamics of only one mode $h^{\alpha s}_{nlm}$ with $(s,n,l,m)=(s_{\alpha},n_{\alpha},l_{\alpha},m_{\alpha})$, which we denote by $h_{\alpha}$. With this reduction, the system is thus described by $\{a_{\alpha}, h_{\alpha}\}$.

\subsection{Canonical quantization}

Now, we turn attention to quantum cosmology, equipped with \ac{hl} gravity and assuming that the dynamics of the homogeneous and isotropic Universe (including any tensor perturbations) is governed by the wave function of the \ac{wdw} equation. From the \ac{hl} action (\cref{eq:HL3}), we can compute the corresponding Hamiltonian and obtain
\begin{subequations}
    \label{eq:hamiltonian}
    \begin{align}
        H & =\sum_{\alpha=1}\left(\Pi_{a_{\alpha}}\dot{a}_{\alpha}+
            \Pi_{h_{\alpha}}\dot{h}_{\alpha}\right)- \mathcal{L} \\
        \begin{split}
            &=\sum_{\alpha=1}
                {\cal V}\left(\frac{N}{a_{\alpha}^{d-2}}\right)\biggl[
                -\frac{1}{2\gamma}\Pi_{a_{\alpha}}^2
                -\frac{\alpha_d}{a^{2}_{\alpha}}\cdots
                -\frac{\alpha_3}{a^{8-2d}_{\alpha}}-\frac{\alpha_2}{a^{6-2d}_{\alpha}}-c_{\rm g}^2
                \frac{d(d-1)}{2a^{4-2d}_{\alpha}}+\frac{\Lambda}{a^{2-2d}_{\alpha}}\\
            &\quad\quad+\frac{2}{{\cal V}^2a^2_{\alpha}}\Pi_{h_{\alpha}}^2
                -\left(\frac{\beta_1}{a^{4-2d}_{\alpha}}
                +\frac{\beta_2}{a^{6-2d}_{\alpha}}+\frac{\beta_3}{a^{8-2d}_{\alpha}}\cdots
                +\frac{\beta_d}{a^2_{\alpha}}\right)h_{\alpha}^2
                \biggr] \, ,
        \end{split}
    \end{align}
\end{subequations}
where $\mathcal{L}$ is defined by $S^{(0)}+S^{(2)} = \int \dd t \mathcal{L}$, $\gamma=d\left(d\lambda-1\right){\cal V}^2$ and the canonical momenta $\Pi_{a_{\alpha}}$ and $\Pi_{h_{\alpha}}$ conjugate respectively to $a_{\alpha}$ and $h_{\alpha}$ are given by
\begin{equation}
    \label{eq:canonicalmomentum}
    \Pi_{a_{\alpha}}
        =\frac{\partial \mathcal{L} }{\partial \dot{a}_\alpha}
        =-d\left(d\lambda-1\right){\cal V} \frac{a_\alpha^{d-2}}{N}\dot{a}_\alpha
    \,, \quad
    \Pi_{h_{\alpha}}
        =\frac{\partial \mathcal{L}}{\partial \dot{h}_\alpha}
        =\frac{{\cal V}}{4} \frac{a_\alpha^d}{N}\dot{h}_\alpha
    \, .
\end{equation}

We can now carry out the canonical quantization of \ac{hl} gravity. This is done by transforming the canonical variables of Hamiltonian mechanics into Hermitian operators that satisfy the canonical commutation relations. The algebra generated by commutative quantities ($c$-numbers) becomes an algebra generated by non-commutative quantities ($q$-numbers) in quantum mechanics. Therefore, an ambiguity in the operator ordering here arises. In performing the canonical quantization, the canonical conjugate momenta are transformed into Hermitian operators, \ie.
\begin{equation}
    \begin{array}{r}
        \Pi_{a_{\alpha}}\mapsto -i\dfrac{\partial}{\partial a_{\alpha}} \\[10pt]
        \Pi_{h_{\alpha}}\mapsto -i\dfrac{\partial}{\partial h_{\alpha}} 
    \end{array}
        \quad 
        \Longrightarrow \quad 
    \begin{array}{l}
        \Pi_{a_{\alpha}}^{2}=
        -\dfrac{1}{a_{\alpha}^{p}}\dfrac{\partial}{\partial a_{\alpha}}\left(a_{\alpha}^{p}
        \dfrac{\partial}{\partial a_{\alpha}}\right) \\[10pt]
        \Pi_{h_{\alpha}}^{2}=
        -\dfrac{\partial^2}{\partial h_{\alpha}^2}
    \end{array}
    \, ,
\end{equation}
where we should take into account the operator ordering ambiguity of $a_{\alpha}$~\cite{Hartle:1983ai}. In quantum cosmology, there exists two well-known orderings: the Laplace-Beltrami operator ordering ($p=1$) and the Vilenkin ordering ($p=-1$) \cite{Steigl:2005fk}.

Once the canonical quantization of the Hamiltonian of \cref{eq:hamiltonian} is realized, we obtain the \ac{wdw} equation
\begin{equation}
    \label{eq:WDW_equation_with_old_coefficients}
    \begin{aligned}
        \sum_{\alpha}\frac{1}{a_{\alpha}^{d-2}}\, \Biggl\{ &
            \frac{1}{2\gamma}\left(\frac{\partial^2}{\partial a^2_{\alpha}} + \frac{p}{a_{\alpha}}\frac{\partial}{\partial a_{\alpha}}\right) -\frac{\alpha_d}{a^{2}_{\alpha}} - \cdots -\frac{\alpha_2}{a^{6-2d}_{\alpha}}-c_{\rm g}^2 \frac{d(d-1)}{2a^{4-2d}_{\alpha}}+\frac{\Lambda}{a^{2-2d}_{\alpha}} \\
        &
            -\frac{2}{{\cal V}^2a^2_{\alpha}}\frac{\partial^2}{\partial h_{\alpha}^2} -\left(\frac{\beta_1}{a^{4-2d}_{\alpha}} +\frac{\beta_2}{a^{6-2d}_{\alpha}}+\cdots +\frac{\beta_d}{a^2_{\alpha}}\right)h_{\alpha}^2\Biggr\} \Psi\left(\{a_{\alpha}, h_{\alpha}\}\right)=0 \, ,
    \end{aligned}
\end{equation}
where $\Psi\left(\{a_{\alpha}, h_{\alpha}\}\right)$ is the wave function of the entire Universe.
By separating the variables, one can easily obtain special solutions of the form $\Psi = \prod_{\alpha}\Psi_{\alpha}(a_{\alpha}, h_{\alpha}; C_{\alpha})$, where $\{C_{\alpha}\}$ are separation constants satisfying $\sum_{\alpha}C_{\alpha}=0$. 
A general solution can be then written as a linear combination of these special solutions, explicitly
\begin{equation}
    \Psi\left(
    \{a_{\alpha}, h_{\alpha}\}\right)=\int \Biggl(\prod_{\beta}\dd C_{\beta}\Biggr)
    A_{\{C_{\beta}\}}
    \prod_{\alpha}\Psi_{\alpha}\left(a_{\alpha}, h_{\alpha}; C_{\alpha}\right)\,.
\end{equation}
where each $\Psi_{\alpha}\left(a_{\alpha}, h_{\alpha}; C_{\alpha}\right)$ represents the wave function of one local Universe $\Sigma_{\alpha}$.

We now redefine the variables and constants in \cref{eq:WDW_equation_with_old_coefficients} as well as the separation constant $C_{\alpha}$. 
\begin{equation*}
    \begin{gathered}
        \mathfrak{h}= \frac{h_{\alpha}}{2\sqrt{\gamma}}\,;
            \qquad
            \mathcal{C} = \gamma C_{\alpha}\,; \\ 
        g_d = \gamma\alpha_d \,,\quad \cdots\,, \quad
            g_2 = \gamma\alpha_2 \,,\quad
            g_1 = \frac{\gamma c_{\rm g}^2d(d-1)}{2} \,,\quad
            g_0 = \gamma \Lambda \,; \\
        f_d = -8\gamma^2 \beta_d \,, \quad \cdots\,, \quad
            f_2 = -8\gamma^2 \beta_2 \,, \quad
            f_1 = -8\gamma^2 \beta_1 \,;
    \end{gathered}
\end{equation*}
The WDW equation for each local universe can then be re-written as 
\begin{align}
    \label{eqn:WDWeq}
    & \left\{
    \frac{1}{2}\left(\frac{\partial^2}{\partial a^2} + \frac{p}{a}\frac{\partial}{\partial a}\right)
    + \left( \mathcal{C}a^{d-2} - \frac{g_d}{a^2} - \cdots
    -\frac{g_{2}}{a^{6-2d}}-\frac{g_{1}}{a^{4-2d}} +\frac{g_{0}}{a^{2-2d}}\right)\right. \nonumber\\
    &\quad\quad\quad \left.
    -\frac{1}{2{\cal V}^2a^2}\frac{\partial^2}{\partial\mathfrak{h}^2} 
    + \frac{\mathfrak{h}^2}{2}\left( \frac{f_1}{a^{4-2d}}+\frac{f_2}{a^{6-2d}}+
    \cdots+ \frac{f_d}{a^2}\right) \right\}
    \Psi(a,\mathfrak{h}) = 0 \, ,
\end{align}
where we have dropped the index $\alpha$ and abbreviated $\Psi_{\alpha}\left(a_{\alpha}, h_{\alpha}; C_{\alpha}\right)$ as $\Psi(a,\mathfrak{h})$.

\section{DeWitt wave function with tensor perturbations}
\label{sec:with-tensor} 

In this section, we compute the $d+1$-dimensional DeWitt wave function for a universe represented by a homogeneous and isotropic background and tensor perturbation around it. In particular we show that the DeWitt wave function near the classical big-bang singularity features scale-invariant tensor perturbations due to the anisotropic scaling with $z=d$.

\subsection{Analytical estimation}
\label{sec:analytical-estimation} 

We will start by extending the results of our previous work~\cite{Matsui:2021yte} to $d + 1$ dimensions. In order to obtain a solution of the WDW equation (\cref{eqn:WDWeq}) that satisfies the DeWitt boundary condition, \ie. $\Psi(0,\mathfrak{h}) = 0$ for all $\mathfrak{h}$, we adopt the following expansion around $a=0$,
\begin{equation}
    \label{eqn:Psi-expanded}
    \Psi(a,\mathfrak{h}) = a^c\sum_{i=0}^{\infty} F_i(\mathfrak{h})\, a^i\,,
\end{equation}
where $F_0(\mathfrak{h})$ is not identically zero and $c$ is a positive constant to be determined. For $\Psi(a,\mathfrak{h})$ to be normalizable on constant-$a$ hypersurfaces, we also impose the following condition on $\Psi(a,\mathfrak{h})$
\begin{equation}
    \label{eqn:bc-phi}
    \lim_{\mathfrak{h}\to\pm\infty} F_i(\mathfrak{h}) = 0 \,.
\end{equation}
If this condition (\cref{eqn:bc-phi}) is not satisfied, the correlation functions such as the power spectrum of $\mathfrak{h}$ on constant-$a$ hypersurfaces would diverge. In order to determine the positive constant $c$, we use the above mentioned requirement that $F_0(\mathfrak{h})$ be a non-trivial smooth function satisfying the condition of \cref{eqn:bc-phi} (with $i=0$).

In \ac{gr}, we previously demonstrated that there are no parameter set for which $F_0(\mathfrak{h})$ can satisfy the condition of \cref{eqn:bc-phi} (with $i=0$) and the tensor perturbations are not smoothly suppressed~\cite{Matsui:2021yte}. In other words, no DeWitt wave function is normalizable on constant-$a$ hypersurfaces around the classical big-bang singularity, unless higher spatial derivative terms are present, as we will see in the following.

Next, let us turn our attention to \ac{hl} gravity. By substituting the small-$a$ expansion (\cref{eqn:Psi-expanded}) into the \ac{wdw} equation (\cref{eqn:WDWeq}), and solving it order by order, we can explicitly determine the functions $F_i$. For $i=0,1,2$, we find the following three equations
\begin{subequations}
    \begin{align}
        \partial_{\mathfrak{h}}^2F_0 - {\cal V}^2\left[ f_d\mathfrak{h}^2 + (c+p-1)c - 2 g_d \right]F_0 & = 0\,, 
        \label{eqn:F0}\\
        \partial_{\mathfrak{h}}^2F_1 - {\cal V}^2\left[ f_d\mathfrak{h}^2 + (c+p)(c+1) - 2 g_d \right]F_1 & = 0\,, \label{eqn:F1}\\
        \partial_{\mathfrak{h}}^2F_2 - {\cal V}^2\left[ f_d\mathfrak{h}^2 + (c+p+1)(c+2) - 2 g_d \right]F_2 & = {\cal V}^2(f_d\mathfrak{h}^2-3g_{d-1})F_0\,. \label{eqn:F2}
    \end{align}
\end{subequations}

Let us first consider the first relation, \ie. \cref{eqn:F0}.
By requiring (\ref{eqn:bc-phi}) (with $i=0$) and the continuity of both $F_0(\mathfrak{h})$ and $\partial_{\mathfrak{h}}F_0(\mathfrak{h})$ at $\mathfrak{h}=0$, we obtain
\begin{equation}\label{eqn:solF0}
 F_0(\mathfrak{h}) = \left\{ \begin{array}{ll}
 \frac{A}{\sqrt{\mathfrak{h}}} W_{N+1/4,1/4}(w)\,, & \ (\mathfrak{h}>0)\\
 -\frac{A}{\sqrt{-\mathfrak{h}}} W_{N+1/4,1/4}(w)\,, & \ (\mathfrak{h}<0)
\end{array}\right.\,,
\end{equation}
where $W_{\mu, \nu}(w)$ is the Whittaker function, $A$ 
is some constant, $N$ some non-negative integer, and $w$ and $c$ are defined by
\begin{equation}\label{eqn:eq-c}
    w = {\cal V}\sqrt{f_d}\mathfrak{h}^2\,, \quad
    c^2 + (p-1)c + \left[\frac{\sqrt{f_d}}{{\cal V}}(4N+1) - 2g_d\right] = 0 \,,
\end{equation}
which, in turn, constrains $f_d$ to be strictly positive.

For \cref{eqn:F1}, by again requiring (\ref{eqn:bc-phi}) (with $i=1$) and the continuity of both $F_1(\mathfrak{h})$ and $\partial_{\mathfrak{h}}F_1(\mathfrak{h})$ at $\mathfrak{h}=0$, and refraining from fine-tuning, we find $F_1(\mathfrak{h}) \equiv 0$ as the next-to-leading order solution in $a$.
However, at the next-to-next-to-leading order, $F_2$ cannot be identically zero.
Once $F_0(\mathfrak{h})$ is given, $F_2$ is determined by \cref{eqn:F2,eqn:bc-phi} (with $i=2$).
While the computation is straightforward, the explicit expression for $F_2$ is fairly long. Hence, we show only its structure, for $N=0,1,2$, without explicit expressions. Using the leading-order solution for different values of $N$, \ie.
\begin{equation}
    \label{eq:F0_for_different_N}
    F_0(\mathfrak{h}) =
    \begin{cases}
        A_0 \exp\left(-\frac{1}{2}w\right)\,, & (N=0)\\
        (1-2w) A_1 \exp\left(-\frac{1}{2}w\right)\,, & (N=1)\\
        \left(1-4w + \frac{4}{3}w^2\right) A_2 \exp\left(-\frac{1}{2}w\right)\,, & (N=2)
    \end{cases}
    \,,
\end{equation}
where $A_N$ ($N=0,1,2$) is some integration constant, the solution for $F_2$ takes the form
\begin{equation}
    \label{eq:F2}
 F_2(\mathfrak{h}) = \sum_{\tilde{N}=0}^{N+1}a_{N,\tilde{N}}w^{\tilde{N}}A_N\exp\left(-\frac{1}{2}w\right)\,.
\end{equation}
Here, $a_{N,\tilde{N}}$ ($\tilde{N}=0,\cdots,N+1$) are constant coefficients determined by the parameters of the \ac{wdw} equation (\cref{eqn:WDWeq}).

If $N=0$, we obtain the ground state of the DeWitt wave function, which, in the $a\to +0$ limit, can be expressed as 
\begin{equation}
 \Psi(a,\mathfrak{h}) = A({\cal V}\sqrt{f_d})^{1/4} a^c \left[ e^{-\frac{{\cal V}\sqrt{f_d}\mathfrak{h}^2}{2}} + \mathcal{O}(a^2)\right]\,.
\end{equation}
Since $f_d\propto n^{2d}$ for large $n$, the corresponding correlation function for the tensor perturbations then reads
\begin{equation}
 \lim_{a\to+0} \langle \mathfrak{h}^2\rangle = 
  \lim_{a\to+0} \frac{\int \dd\mathfrak{h}\, \mathfrak{h}^2 \left|\Psi(a,\mathfrak{h})\right|^2 }
    {\int \dd\mathfrak{h}\,  \left|\Psi(a,\mathfrak{h})\right|^2}
    = \frac{1}{2{\cal V}\sqrt{f_d}} \propto n^{-d}\,, \quad \mbox{for large }\, n\,\mbox{ and }\, N=0\,.
\end{equation} 
This clearly shows the scale-invariance and the finiteness of the power spectrum of $\mathfrak{h}$ for $f_d>0$ (see \cite{Matsui:2021yte} for the $d=3$ case). Also in the general case of \cref{eqn:solF0} with general $N$ ($\geq 0$), the power spectrum and other correlation functions of $\mathfrak{h}$ are scale-invariant and finite in the UV limit ($a\to+0$) as far as $f_d>0$. In particular, 
\begin{equation}
 \mathcal{M}_{\rm HL}^{3d-3} \lim_{a\to+0} n^d \langle \mathfrak{h}^2\rangle = \mathcal{O}(1)\times {\cal M}^{d-1}\,,\quad \mbox{for large }\, n\,\mbox{ and any }\, N\geq 0\,,
\end{equation}
where $\mathcal{M}$ is the mass scale such that $\beta_d \sim n^{2d}/\mathcal{M}^{2d-2}$ (and thus ${\cal V}\sqrt{f_d}\sim n^d\mathcal{M}_{\rm HL}^{3d-3}/\mathcal{M}^{d-1}$) for large $n$. This is completely consistent with the result of \cite{Mukohyama:2009gg}.

We have seen that \cref{eqn:eq-c} and eqs.~\eqref{eqn:F2} and \eqref{eqn:bc-phi} (with $i=0$) requires $f_d>0$. Therefore, if $f_d\leq 0$ then there is no non-trivial smooth solution satisfying \cref{eqn:bc-phi} (with $i=0$). In particular, this is the case in the absence of $z=d$ terms (for which $f_d=0$). This no-go result applies to many gravitational theories (including \ac{gr}) for which the action does not contain terms with $2d$ spatial derivatives.

\subsection{Numerical solution}
\label{sec:numerical-estimation} 

\begin{figure}
    \centering
    \input{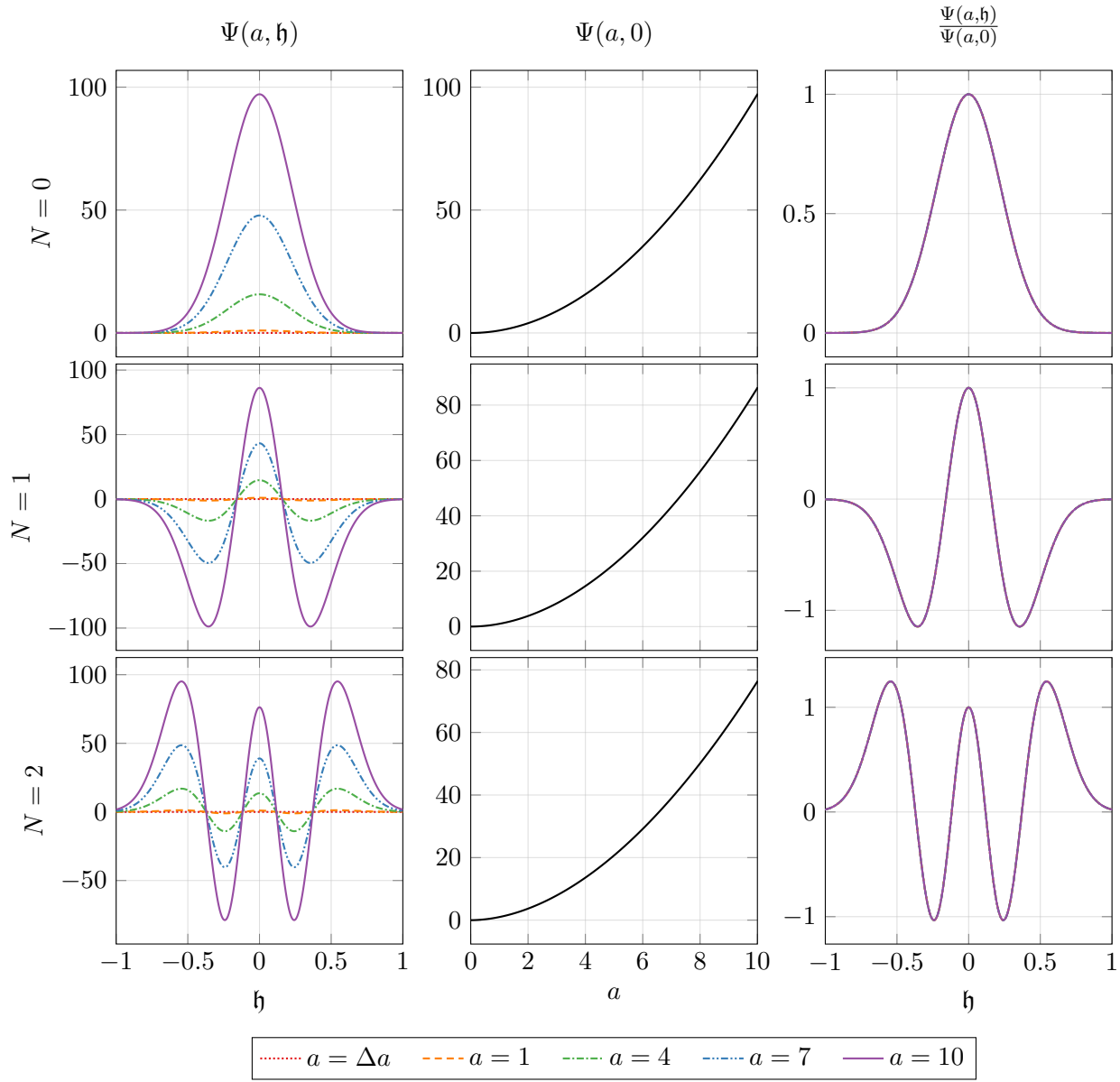}
    \caption{
        Numerical integration result for the parameter set $f_2=0$, $\mathcal{C}=0$, $g_0=0$.
        For any $N$, while the amplitude of $\Psi$ here exponentially increases (central column) with $a$, the initial shape is conserved unchanged (right column).
    }
    \label{fig:beta2=0_C=0_Lam=0}
\end{figure}

\begin{figure}
    \centering
    \input{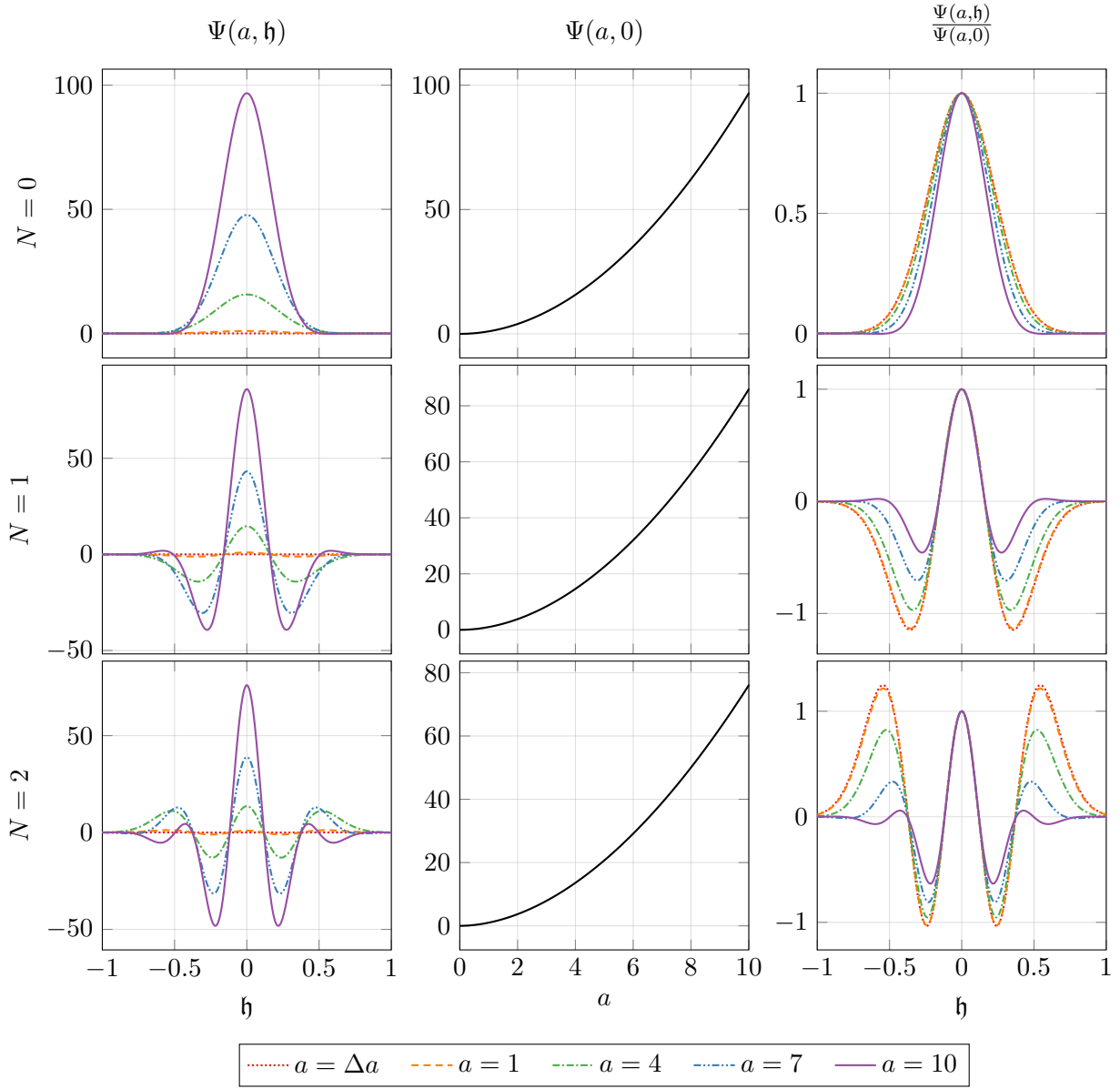}
    \caption{
        Numerical integration result for the parameter set $f_2=1$, $\mathcal{C}=0$, $g_0=0$.
        While overall similar, unlike the case $f_2 = \mathcal{C} = g_0 = 0$ of \cref{fig:beta2=0_C=0_Lam=0}, the shape of the wave function evolves significantly through the ``evolution'' towards larger $a$, as the rightmost columns exhibits.
        Essentially, the amplitude's growth is slowed down the further it is from $\mathfrak{h}=\num{0}$ (central and right columns) as $f_2$ actually enables a term in $\mathfrak{h}^2$ in \cref{eq:WDW_equation_with_numerical_values}.
    }
    \label{fig:beta2=1_C=0_Lam=0}
\end{figure}

\begin{figure}
  \centering
  \input{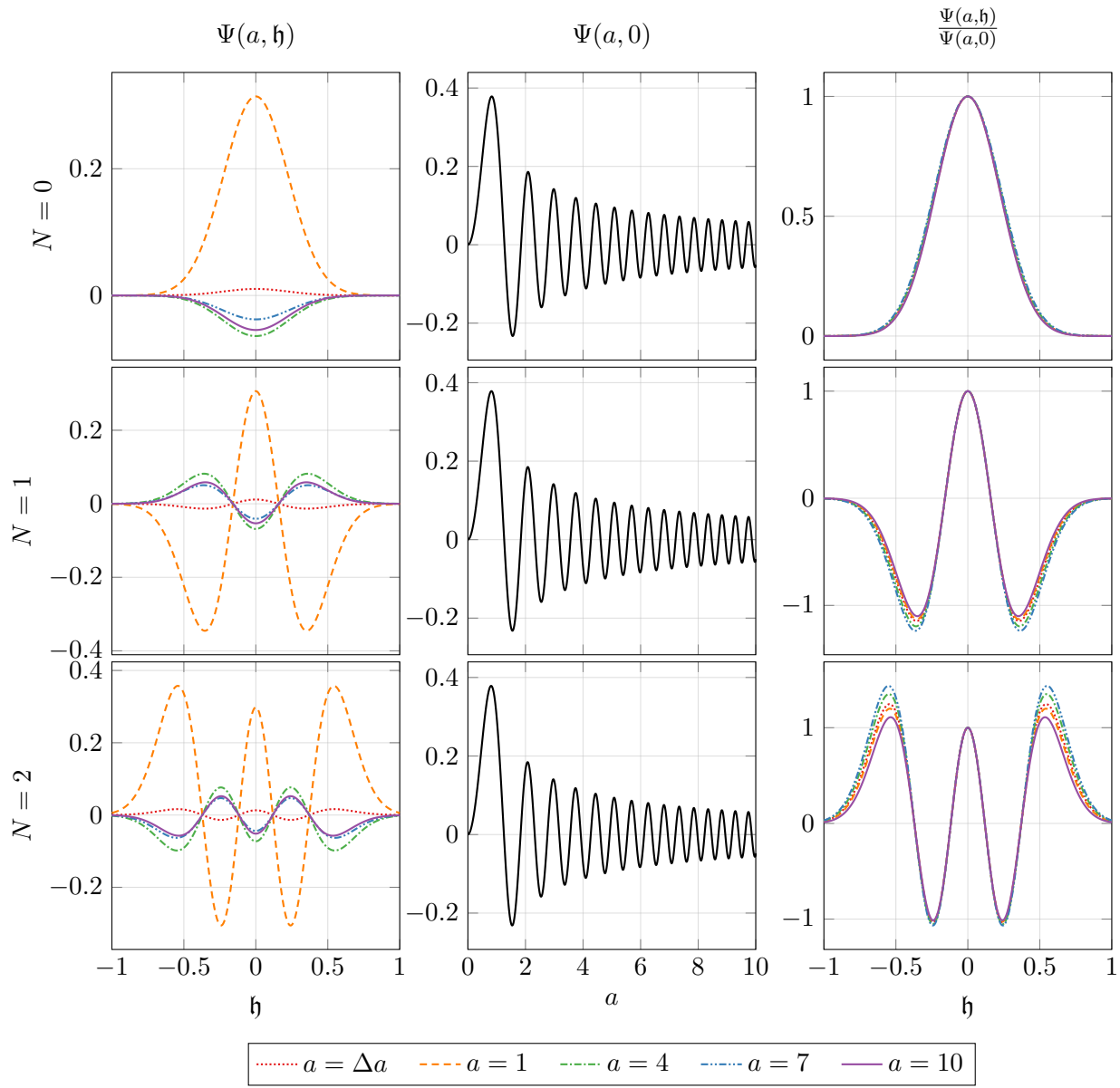}
  \caption{
      Numerical integration result for the parameter set $f_2=1$, $\mathcal{C}=10$, $g_0=0$.
      Now that $\mathcal{C}$ is set to $10$, a term in $a$ now enters \cref{eq:WDW_equation_with_numerical_values} and will quickly dominate over the previous $-a^{-2}$ term.
      In so doing, it brings the system in a regime of damped oscillation (central column).
      The effect of $f_2$ (see \cref{fig:beta2=1_C=0_Lam=0}) remains, though now slightly muted (right column).
  }
  \label{fig:beta2=1_C=10_Lam=0}
\end{figure}

\begin{figure}
    \centering
    \input{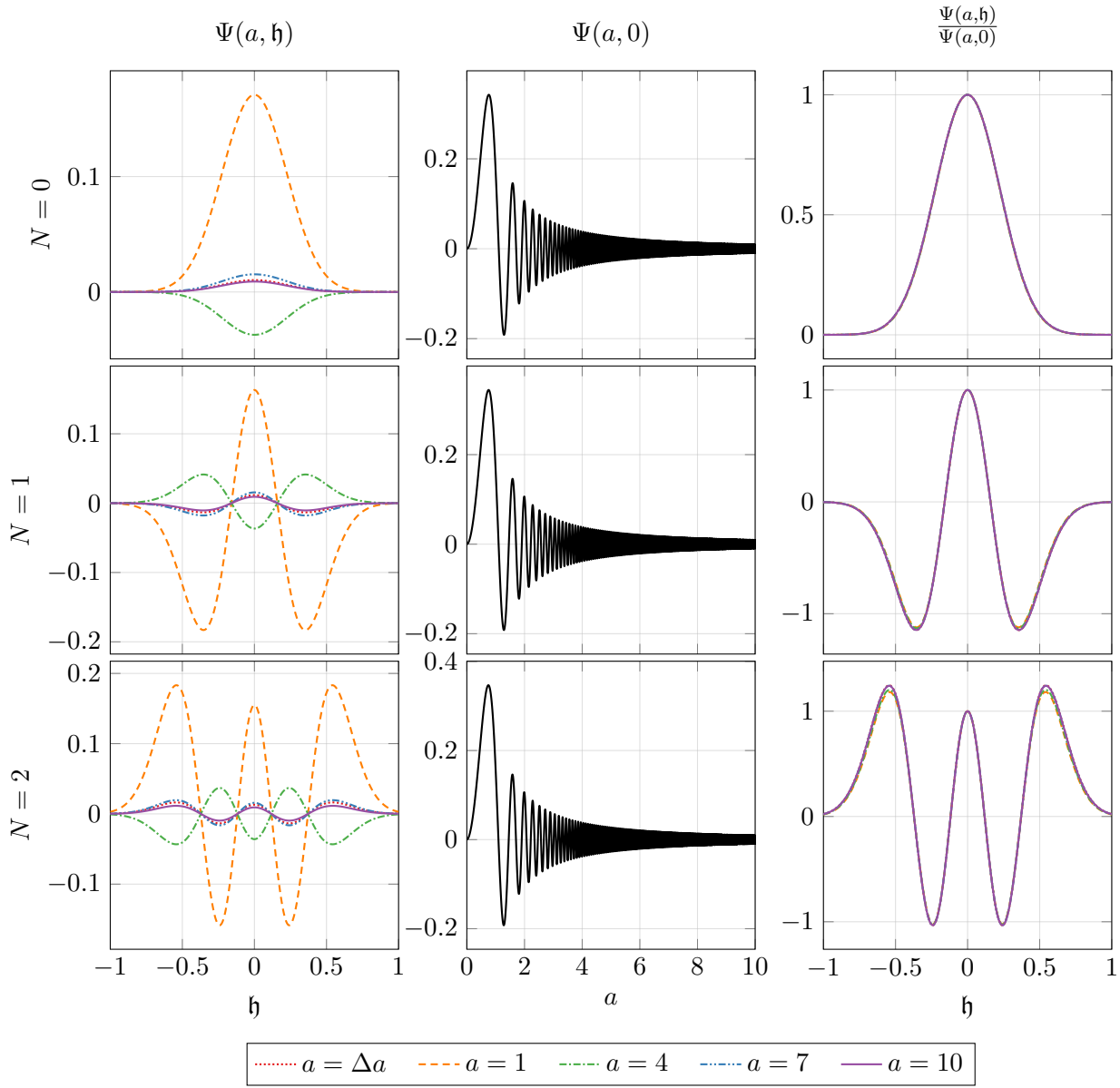}
    \caption{
        Numerical integration result for the parameter set $f_2=1$, $\mathcal{C}=10$, $g_0=10$.
        The $g_0$ coefficient now brings a $a^4$ term, that will eventually overcome the previous $a$ and $-a^{-2}$ contributions, pushing the system into an even more strongly damped oscillation regime (central column).
        The influence of $f_2$ remains visible, though now barely visible (right column).
    }
    \label{fig:beta2=1_C=10_Lam=10}
\end{figure}

Numerically integrating the \ac{wdw} equation (\cref{eqn:WDWeq}) is fairly straightforward. The equation is hyperbolic, and one can use the expansion around $a=0$ obtained in the previous~\cref{sec:analytical-estimation} as the initial condition on a constant-$a$ hypersurface, provided that the initial value of $a$ is sufficiently small. The initial condition is uniquely specified by the non-negative integer $N$ up to an overall amplitude. For concreteness, we fix the overall amplitude by 
\begin{equation}
    \lim_{a\to +0} a^{-c}\Psi(a,0) = 1\,.
\end{equation}
This translates into $A_N=1$ for all $N=0,1,2$.

In the numerical study, we shall only consider the $3+1$ dimensional case, employ the second-order finite difference scheme for derivatives with respect to $\mathfrak{h}$, and adopt two different integration methods to evolve the system with respect to $a$. Both the iterated Crank-Nicolson method with two iterations and the fourth-order Runge-Kutta method have shown identical satisfactory results~\footnote{The two independent codes were previously used in \cite{Berti:2022hwx} and one of them was used also in \cite{Bhattacharjee:2018nus,Mukohyama:2020lsu}.}. To this end, the \enquote{space} $\mathfrak{h}$ is sampled on a grid of \num{40000} points spaced by $\Delta\mathfrak{h}=\num{1e-4}$. As for the \enquote{time}, the range in $a$ is discretized into $10^6$ steps, each of size $\Delta a = 10^{-1}\cdot\Delta\mathfrak{h} = \num{1e-5}$. We start from the small initial \enquote{time} $a=a_0$ by using the solution obtained in~\ref{sec:analytical-estimation}\,---the expansion of \cref{eqn:Psi-expanded} with \cref{eq:F0_for_different_N,eq:F2}---, which is valid for sufficiently small $a_0$. Here, we choose $a_0 = 10^{-1}\cdot\Delta a = \num{1e-6}$.

For simplicity we set $p=1$, $g_3=2$, $f_3=1$, and $g_2$, $g_1$, $f_1$ and $f_0$ to $0$ (with $d=3$) in all simulations shown hereafter. Then, the Wheeler-DeWitt equation (\ref{eqn:WDWeq}) is specified by the remaining parameters $f_2$, $\mathcal{C}$ and $g_0$. It can now be compactly written as
\begin{equation}
    \label{eq:WDW_equation_with_numerical_values}
    \left\{
        \frac{1}{2}
        \left(
            \frac{\partial^2}{\partial a^2}
            +
            \frac{1}{a} \frac{\partial}{\partial a}
        \right)
        +
        \left[
            \mathcal{C} a
            - \frac{2}{a^2}
            + g_0 a^4
            + \left(f_2 + \frac{1}{a^2}\right)\frac{\mathfrak{h}^2}{2}
        \right]
        - \frac{1}{2 {\cal V}^2 a^2} \frac{\partial^2}{\partial \mathfrak{h}^2}
    \right\}
    \Psi (a, \mathfrak{h})
    = 0\,.
\end{equation}
From thereon, we show the numerical results in \cref{fig:beta2=0_C=0_Lam=0,fig:beta2=1_C=0_Lam=0,fig:beta2=1_C=10_Lam=0,fig:beta2=1_C=10_Lam=10}, and make some comments.

In the four parameter sets exhibited here, we turn on one after the other each of the three parameters we left free, \ie. $f_2$, $\mathcal{C}$ and $g_0$. In \cref{fig:beta2=0_C=0_Lam=0}, we set $f_0=\mathcal{C}=g_0=0$ and there are only two non-derivative terms, both of which are proportional to $a^{-2}$. The system in this case exhibits the exact $z=3$ anisotropic scaling. The initial input here simply exponentially grows, without its form being altered. This is easily understood by observing the separability of \cref{eq:WDW_equation_with_numerical_values} with respect to $\mathfrak{h}$ and $a$. For $f_2=1$, on the other hand, a term $\propto \mathfrak{h}^2\cdot a^0$ enters and, as seen in \cref{fig:beta2=1_C=0_Lam=0}, visibly modifies the shape of the wave function.

In the next and last two figures, \cref{fig:beta2=1_C=10_Lam=0,fig:beta2=1_C=10_Lam=10}, as we set $\mathcal{C}$ then $g_0$ to 10, we enable, in turn, terms $\propto a$ and $\propto a^4$, respectively, in \cref{eq:WDW_equation_with_numerical_values}. They will drive the system into a damped oscillation regime; $g_0 a^4$ more strongly so than $\mathcal{C} a$. As they come to dominate, the effects of $f_2 \mathfrak{h}^2$ and $\mathfrak{h}^2/a^2$, while remaining, are proportionately muted.

Overall, the afore-described behaviors are as expected from \cref{eq:WDW_equation_with_numerical_values} (or \cref{eqn:WDWeq}). Most importantly, the wave function is obviously well-defined and normalizable on constant-$a$ hypersurfaces, all the way from the classical big-bang singularity to finite values of $a$.

\section{Conclusion}
\label{sec:conclusion} 

We have investigated the behavior of a $d+1$-dimensional Universe near the classical big-bang singularity based on the Wheeler-DeWitt equation and the DeWitt boundary condition, which amounts to a vanishing wave function of the Universe at the classical singularity. For concreteness we have studied a homogeneous and isotropic background and tensor perturbation around it described respectively by the scale factor $a$ and the amplitude of tensor perturbation $\mathfrak{h}$ as a simple model of the Universe. In general relativity, the DeWitt wave function for $\mathfrak{h}$ on constant-$a$ hypersurfaces is not normalizable near $a=0$, meaning that the perturbative expansion breaks down. On the contrary, in Ho\v{r}ava-Lifshitz gravity the higher dimensional operators required by the perturbative renormalizability of the theory render the wave function of the tensor perturbation normalizable, all the way from the classical big-bang singularity at $a=0$ to finite values of $a$. The DeWitt wave function for $\mathfrak{h}$ on constant-$a$ hypersurfaces is of a Gaussian shape for the ground state and a Gaussian multiplied by an even polynomial of $\mathfrak{h}$ for excited states. We have analytically proved these behaviors of the wave function near $a=0$ for any $d$, and numerically demonstrated them in a finite interval from $a=0$ to a finite value of $a$. These results provide strong evidences for close connections between the regularity of the DeWitt wave function of the Universe and the renormalizability of HL gravity.

\acknowledgments

Two of us (H.M. and S.M.) would like to thank Atsushi Naruko for collaborating on the previous work \cite{Matsui:2021yte} on the subject.
P.M. acknowledges support from the Japanese Government (MEXT) scholarship for Research Student.
The work of H.M. was supported by JSPS KAKENHI Grant No. JP22J01284.
S.M.'s work was supported in part by JSPS Grants-in-Aid for Scientific Research No.~17H02890, No.~17H06359, and by World Premier International Research Center Initiative, MEXT, Japan.

\appendix
\section{Ho\v{r}ava-Lifshitz DeWitt wave function without tensor perturbations}
\label{sec:without-tensor} 

We review quantum cosmology based on the \ac{hl} gravity and consider the DeWitt wave function in the \ac{wdw} equation (\cref{eqn:WDWeq}) for a homogeneous and isotropic universe without any perturbations. For only $a$ in $3+1$ dimensions, the \ac{hl} DeWitt wave function has already been investigated in previous works \eg. Refs.~\cite{Bertolami:2011ka,Ali:2011sv,Vakili:2013wc}. However, the \ac{hl} gravity introduces an extra component that behaves like a pressure-less dust and that is called dark matter as an integration constant~\cite{Mukohyama:2009mz,Mukohyama:2009tp}. The new dark matter component was derived in the classical cosmology. Correspondingly, when one considers the \ac{wdw} equation, the \ac{hl} gravity naturally introduces a dark matter as a separation constant (instead of an integration constant) into the local universe~\cite{Matsui:2021yte}. Therefore, the extra component potentially alters the quantum cosmology based on the HL gravity.

Hereafter, we will consider the evolution of the early universe in the \ac{hl} quantum cosmology. In the anisotropic scaling regime, the \ac{hl} term of $g_3$ dominates and the solution of the \ac{wdw} equation (\cref{eqn:WDWeq} with $\partial/\partial h$ and $h$ removed) approximately takes the from $\Psi(a)=\mathcal{A}_1a^{\tilde{c}}$, where $\mathcal{A}_1$ is a normalization constant~\cite{Bertolami:2011ka} and $\tilde{c}$ is a constant determined by $\tilde{c}(\tilde{c}-1)+p\tilde{c}-2g_3=0$, \ie. 
\begin{equation}\label{exponent-z}
 \tilde{c} = \frac{1}{2} \left(1-p+\sqrt{(1-p)^2+8g_3}\right)\,,
\end{equation}
where we must impose $g_3>-(1-p)^2/8$ to ensure the regularity of the solution. Recalling that the DeWitt wave function is regular at the classical big-bang singularity, the wave function describes a universe that has emerged from the initial singularity to the anisotropic scaling regime.

In order to highlight the effects of the dark matter as an integration constant (or as a separation constant), \ie. the ${\cal C}$ term, we set $g_2=g_1=g_0=0$. In this case the \ac{wdw} equation (\cref{eqn:WDWeq} with $\partial/\partial h$ and $h$ removed) reads,
\begin{align}
\left\{\frac{1}{2}\left(\frac{\partial^2}{\partial a^2} + \frac{p}{a}\frac{\partial}{\partial a}\right)+\mathcal{C}a - \frac{g_3}{a^2}\right\}\Psi\left(a\right)=0\,.
\end{align}
Thus, imposing the DeWitt boundary condition on the wave function and using the asymptotic form of $J_{\nu}(z)$, we can get the following wave function
\begin{align}
    \begin{split}
        &\Psi(a)=\mathcal{A}_2\, 3^{\frac{p-1}{3}} a^{\frac{1-p}{2}} (2 \mathcal{C} )^{\frac{1-p}{6}}  \Gamma \left(1+\frac{1}{3} \sqrt{(1-p)^2+8 g_3}\right)
        J_{\frac{1}{3} \sqrt{(1-p)^2+8 g_3}}\left(\frac{2}{3} a^{3/2} \sqrt{2 \mathcal{C}}\right)\,,
    \end{split}
\end{align}
whose asymptotic forms are given by
\begin{align}
\begin{split}
&\Psi(a\ll 1)\sim  \mathcal{A}_2\, (\mathcal{C}a )^{\frac{1-p+\sqrt{(1-p)^2+8 g_3}}{6}},\\
&\Psi(a \gg 1)\sim  \mathcal{A}_2\, a^{-\frac{1+2p}{4}} (2 \mathcal{C})^{-\frac{1+2p}{12}} 
\cos\left(\frac{2}{3} a^{3/2} \sqrt{2 \mathcal{C}
 }-\frac{1+\frac{2}{3} \sqrt{(1-p)^2+8 g_3}}{4}\pi\right)\,.
\end{split}
\end{align}
Here, $\mathcal{A}_2$ is a normalization constant. These expressions show that the DeWitt wave function is suppressed and oscillates, respectively, in the anisotropic scaling regime and the dark-matter dominated regime.
After the Universe emerges through the classical big-bang singularity and the anisotropic scaling regime, the dark matter as an integration constant (or as a separation constant) represented by the parameter $\mathcal{C}$ dominates the energy density of the universe.
Subsequently, in such a local universe, the dark matter component may eventually decay into ordinary matter and radiation~\cite{Bramberger:2017tid}, and in this case the cosmological evolution would be smoothly connected to the usual big-bang cosmology.

\bibliography{Refs.bib}

\end{document}